# Non-Ising Character of a Ferroelectric Wall Arises from a Flexoelectric Effect

Yijia Gu[1], Menglei Li[1,2], Anna N. Morozovska[3,4], Yi Wang[1], Eugene A. Eliseev[4], Venkatraman Gopalan[1], and Long-Qing Chen[1]

[1] Department of Materials Science and Engineering, Pennsylvania State University, University Park, PA 16802, USA

[2] Department of Physics and State Key Laboratory of Low-Dimensional Quantum Physics, Tsinghua University, Beijing 100084, People's Republic of China

[3] Institute of Physics, National Academy of Science of Ukraine, 46, pr. Nauki, 03028 Kiev, Ukraine

[4] Institute for Problems of Materials Science, National Academy of Science of Ukraine, Krjijanovskogo 3, 03142 Kiev, Ukraine

Abstract

Using the classic ferroelectric BaTiO$_3$ as an example, we show that the 180° ferroelectric domain wall, long considered to be of Ising-type and charge neural, contains both Bloch and Néel type polarization components, and is thus charged. The bound charge density at the wall may reach as high as $10^6 \sim 10^7$ C/m$^3$. It is demonstrated that the flexoelectric effect arising from stress inhomogeneity is responsible for the additional Bloch and Néel polarization components. The magnitudes of these additional components are determined by the competing depolarization and flexoelectric fields.

The coupling between electric polarization ($P_i$) and mechanical deformation strain ($\varepsilon_{ij}$) is a fundamental property of materials. This coupling in first order can be written as,

$$P_i = d_{ijk}\varepsilon_{jk} + \mu_{ijkl}\frac{\partial \varepsilon_{kl}}{\partial x_j} \qquad (i,j,k,l=1,2,3) \qquad (1)$$

where $d_{ijk}$ is third-rank piezoelectric tensor and $\mu_{ijkl}$ is known as the fourth-rank flexoelectric (polarization) tensor. The first term on the right-hand side of Equation (1) describes piezoelectric contribution, the linear coupling of polarization and strain, and is present only in materials that lack inversion symmetry. The second term is the flexoelectric contribution, which is a linear dependence of the polarization on strain gradient, and is present in all materials. For example, $SrTiO_3$ and NaCl are not piezoelectric, but they are flexoelectric. Albeit ubiquitous, the flexoelectric effect is usually ignored. This is because the flexoelectric coefficients, $\mu_{ijkl}$, are very small, typically on the order of nC/m [1]. However, when the system size scales down to nanometer scale, the strain gradients can reach $\sim 10^8$ m$^{-1}$, and thus this effect can become significant or even dominate. The domain walls, which have strain variation and thickness on the nanometer scale, are excellent candidates for displaying significant flexoelectricity.

The antiparallel (180°) domain wall is one of the simplest, and is universally present in all ferroelectrics. The spontaneous polarizations in the neighboring domains are both parallel to the domain wall but along opposite directions. It has long been believed that this type of domain wall is Ising-like and charge-neutral, as shown in Figure 1(a). However, recent theoretical studies have found that they are more complex. First-principles calculations showed that the 180° domain wall of tetragonal $BaTiO_3$ is Ising-like, with fluctuations in polarization component perpendicular to the domain wall that did not appear to be spatially correlated with the wall; they were thus dismissed as artifacts [2]. Also using first-principles theory, Lee *et al*. first showed that 180° domain wall of LiNbO$_3$ and PbTiO$_3$ indeed possessed non-Ising character [3]. The hexagonal LiNbO$_3$ exhibits both Bloch-like and Neel-like polarization components [3]. The antiparallel wall in tetragonal PbTiO$_3$ exhibits an Ising-Néel like polarization configuration [3] (Figure 1(b)) and additional Bloch-Néel-like features by changing the domain wall orientation [4]. A Landau-Devonshire theoretical analysis demonstrated that the 180° domain walls of tetragonal BaTiO$_3$ are bichiral, i.e. two orthogonal polarization components parallel to the wall [5] (Fig 1(c)). From all these theoretical calculations, we conclude that the 180° domain walls are predominantly Ising-like, but mixed with Bloch and/or Néel-like fluctuations. Ising-Bloch like (Fig 1(c)) feature of the domain wall character is attributed to the flexoelectric effect [5]. The existence of Néel feature is still under debate.

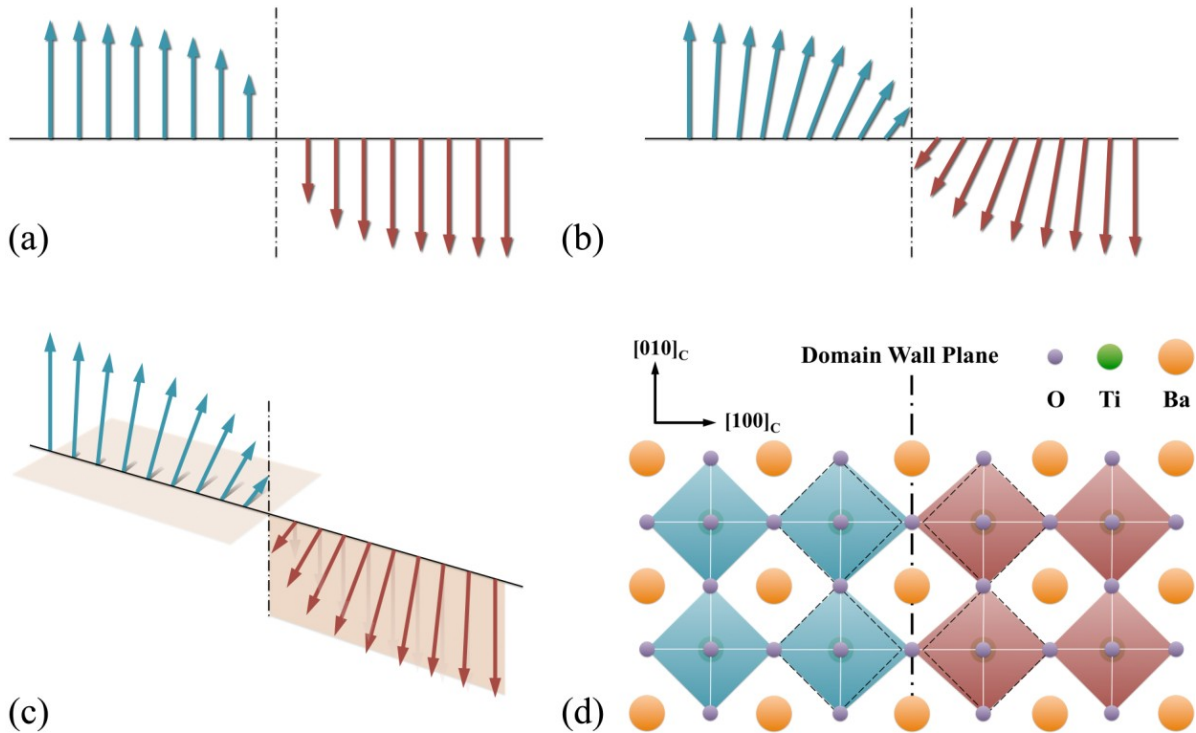

Figure 1. (a) Ising type domain wall. The blue and red arrows represent positive and negative polarizations, respectively. (b) Ising-Néel like domain wall. (c) Ising-Bloch like domain wall. (d) The schematic of 180° domain wall in tetragonal BaTiO$_3$. The blue and red oxygen octahedra indicate positive and negative out-of-plane polarizations, respectively. The adjacent unit cells at the wall are deformed as indicated by the dashed squares.

Assuming the electric polarization comes from the displacement of Ti$^{4+}$ and O$^{2-}$, a 180° domain wall of tetragonal BaTiO$_3$ is sketched in Figure 1(d). The domain wall plane lies on a Ba-O plane, and the neighboring two domains have out-of-plane spontaneous polarizations antiparallel to each other. Due to the mirror symmetry, the oxygen atoms on the wall plane are not displaced, which induces the deformation with respect to the stress-free equilibrium tetragonal unit cell. Since the deformation is confined to the domain wall region that is less than 1nm thick, the strain gradient generated from the domain wall can reach as high as $10^7 \sim 10^8$ m$^{-1}$. Thus, simply from Equation (1), the additional polarization induced by flexoelectric effect is on the magnitude of $10^{-2}$ C/m$^2$, which should be considered (the spontaneous polarization of BaTiO$_3$ at room temperature is 0.26 C/m$^2$). But how significant is the induced polarization, especially the component normal to the wall (Néel feature) which may be greatly weakened or even entirely screened by the depolarization field [5]? And what are the new features of the polarization induced by the flexoelectric effect at the domain walls where bulk symmetry is broken? In this letter, we demonstrate that the 180° domain wall structure of tetragonal perovskite BaTiO$_3$ has **both** Bloch-like and Néel-like features. Both features are found to be strongly anisotropic. Previously reported calculations either missed [3] or ignored [5] one of the two components. By using a combination of phase-field modeling [6,7] and first-principles calculations, we show that the additional Bloch-Néel-like feature is intrinsic to a 180° domain wall and is entirely due to the flexoelectric effect.

We first extended the phase-field model of ferroelectric domains [6,7] to include the flexoelectric contributions. The Gibbs free energy of a ferroelectric crystal is given by [8]

$$G = \alpha_{ij} P_i P_j + \alpha_{ijkl} P_i P_j P_k P_l + \alpha_{ijklmn} P_i P_j P_k P_l P_m P_n + \alpha_{ijklmnor} P_i P_j P_k P_l P_m P_n P_o P_r + \frac{1}{2} g_{ijkl} \frac{\partial P_i}{\partial x_j} \frac{\partial P_k}{\partial x_l}$$
$$- \frac{1}{2} s_{ijkl} \sigma_{ij} \sigma_{kl} - Q_{ijkl} \sigma_{ij} P_k P_l + \frac{F_{ijkl}}{2} (\frac{\partial P_k}{\partial x_l} \sigma_{ij} - \frac{\partial \sigma_{ij}}{\partial x_l} P_k) - P_i (E_i + \frac{E_i^d}{2})$$

(2)

where $x_i$ is the $i$-th component of the Cartesian coordinate system, $P_i$ is the polarization component, $\sigma_{ij}$ is the stress component, $E_i$ is the applied electric field, $E_i^d$ is the depolarization field, $\alpha$'s are the dielectric stiffness tensor (only $\alpha_{ij}$ is assumed to be temperature dependent), $g_{ijkl}$ is the gradient energy coefficient, $s_{ijkl}$ is the elastic compliance tensor, $Q_{ijkl}$ is the electrostrictive tensor, and $F_{ijkl}$ is the flexoelectric tensor. The values of the coefficients for BaTiO$_3$ are from the literature [9–12] (also listed in Table S2 of supplementary materials).

We then theoretically study the orientation dependence of a 180° wall. The setup of the system is illustrated in Fig 2(a) schematically, with the angle $\theta$ representing the rotation angle of the domain wall with respect to the crystallographic direction. The domain wall lies in the $x_2$-$x_3$ plane and perpendicular to the $x_1$ direction. The system is then simplified to a one-dimensional problem with the simulation size $4096\Delta x \times 1\Delta x \times 1\Delta x$ using the three-dimensional phase-field model. Periodical boundary condition is imposed along each direction. The stress of each grid point is calculated using Kachaturyan's microelastic theory [13], and the electric depolarization field is obtained by solving Poisson's equation. For the one dimensional case, the depolarization field is simply $-P_1/(\varepsilon_b \varepsilon_0)$, where $\varepsilon_b$ and $\varepsilon_0$ are the dielectric constant of background [12,14] and vacuum permittivity, respectively. We start from a two-domain structure with only spontaneous $+P_3$ and $-P_3$ in each domain as illustrated in Figure 2 (a), and then let the system relax to equilibrium.

To check the existence of the Néel feature, we calculated the polarization profile of 180° domain wall at $\theta = 0$ ($(100)_C$ plane) as shown in Figure 2 (c). In addition to the $P_3$ component, we observed nonzero $P_1$ component perpendicular to the domain walls while the $P_2$ component remains exactly zero after relaxation. We have carefully checked that although the magnitude of $P_1$ is small, it is not the artifact of the numerical calculation. (A detailed analysis of possible sensitivity of polarization component on various parameters was shown in the supplementary materials)

To validate our results with theoretical calculations from atomic level, we also performed first-principles calculations as shown in Figure 2 (d). The computed bulk $BaTiO_3$ structural parameters in tetragonal phase were a=3.9799 Å, c=4.0768 Å, which are very close to the experimental values of a=3.9970, c=4.0314 [15]. We stacked the 5-atom unit cells in the $x_1$ direction to form a supercell consisting $2N \times 1 \times 1$ ($N$=4,5… 8) unit cells and made the half $N$ cells have initial polarizations $P_3$ pointing up and the other half pointing down. But the atoms in the boundary of the domains were kept at centrosymmetric positions to ensure the mirror symmetry, thus the supercell has the similar two-domain structure as the phase-field simulation. The calculations converged well for $N$ >5. Besides the situation where the $(100)_C$ domain wall lies on Ba-O plane as in the phase-field modeling, a 180° wall can also be centered on Ti-O plane. Both previous study [16] and our calculations show that the BaO-centered domain walls are more stable. Therefore, it's reasonable to focus only on the BaO-centered $(100)_C$ domain wall. The local polarizations were calculated using the method of Meyer and Vanderbilt [16]. During the structure relaxation, we fixed the two lattice vectors parallel to the domain wall plane and optimized the third lattice vector normal to the plane, in order to eliminate the influence of the elastic energy. After the structure relaxation under Ba-centered inversion symmetry constraints, $P_1$ component emerged while $P_2$ component remained zero. The spontaneous polarization $P_3$ calculated from first-principles is around 0.31 C/m$^2$ (Fig 2(d)) which is consistent with the 0 K value extrapolated from the phase-field model. Although the width of Néel wall may not be accurate (affected by the short period of computational cell), the peak position of $P_1$ appears at the first unit cell which is in good agreement with phase-field result. Thus the DFT calculations qualitatively agree with phase-field results and confirm that the 180° domain wall at $\theta = 0$ is Ising-Néel-like.

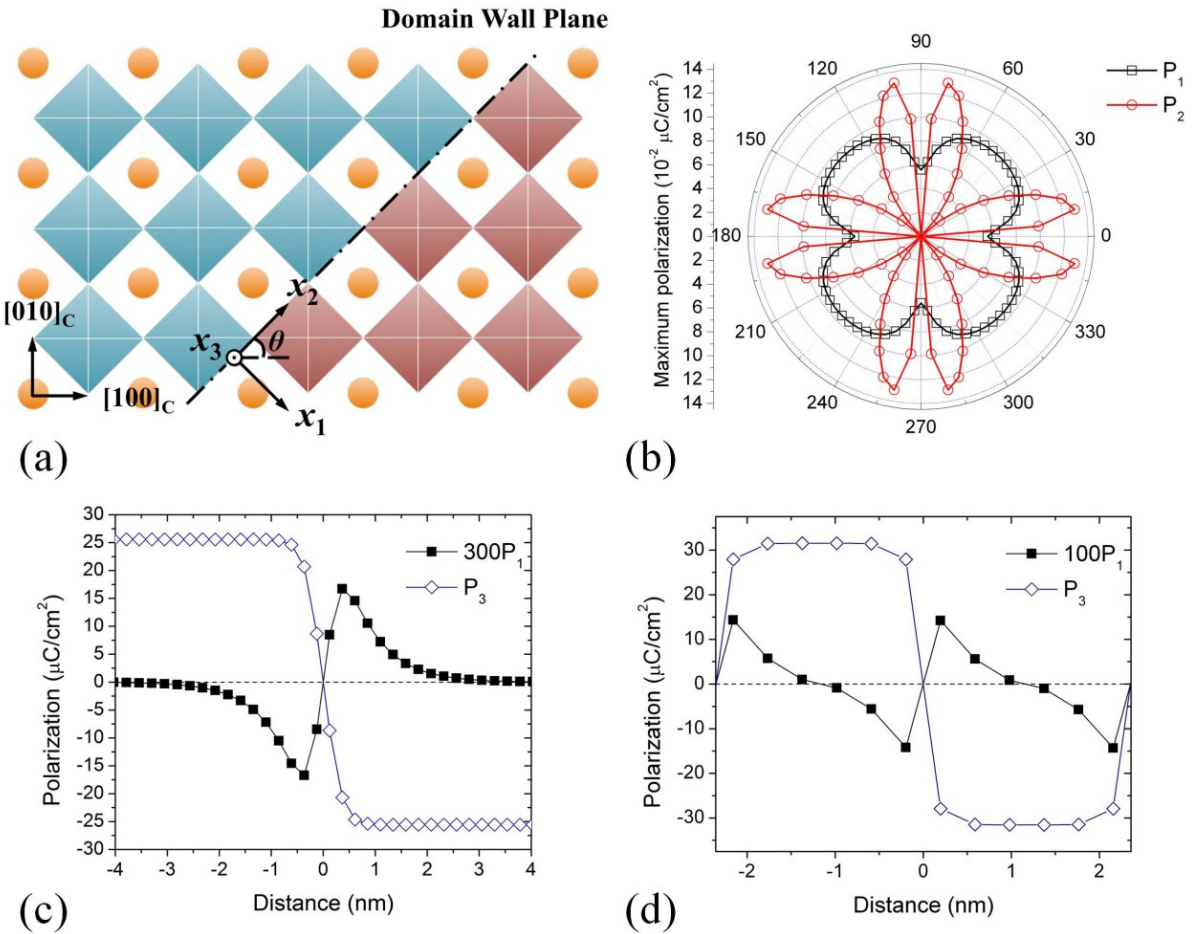

Figure 2. (a) The schematic of system setup for 180° ferroelectric domain wall in tetragonal BaTiO$_3$. Superscript *C* denotes the crystallographic direction. $x_2$ and $x_3$ directions are parallel to the domain wall plane, while $x_1$ is always perpendicular to the wall. $\theta$ indicates the angle between domain wall and the crystallographic direction [100]$_C$. (b) Maximum absolute value of the polarization components induced by the flexoelectric effect in the wall as a function of the rotation angle $\theta$, calculated from phase-field method. (c) The phase-field simulation result of 180° domain wall profile in tetragonal BaTiO$_3$ at $\theta$ = 0. In additon to the conventional Ising wall, a Néel type component is also observed. (d) Polarization profile calculated from first principles calculations. The plot shows a period of the computation cell with two domain walls.

Comparing with the previous model, we believe that the additional flexoelectric contribution leads to the appearance of the Néel-like feature. To convincingly validate our hypothesis, we performed phase-field simulations with the flexoelectric effect turned off. Indeed, at equilibrium, there only exists $P_3$ component in the system, which has the same profile as Figure 2(a). Therefore, we conclude that the appearance of $P_1$ is entirely driven by the flexoelectric effect due to stress inhomogeneity around the domain walls.

Although the additonal polarization components induced by flexoelectricity are small in magnitude, they can significantly impact the behavior of domain walls. As illustrated in Figure 2(c), $P_1$ fluctuates as an odd function through the domain wall, and become zero at the domain wall center. The value of $P_1$ is three orders of magnitude smaller than that of $P_3$. The electric potential change at the domain wall is also weak (order of $10^{-3}$ V). However, the induced bound charge can reach as high as $10^6$ C/m$^3$. Consequently, 180° domain walls will interact with ionic defects such as oxygen vacancies both mechanically and electrostatically. Thus the 180° domain walls may have higher electric conductivity. Actually, J. Guyonnet *et al* [17] have demonstrated that the 180° domain walls in tetragonal Pb(Zr$_{0.2}$Ti$_{0.8}$)O$_3$ thin films have higher conductivity due to charged domain walls.

The domain wall orientation, which affects both the strain gradient and the flexoelectric coefficients, further complicates the character of ferroelectric domain walls. As $\theta$ changes from 0 to $2\pi$, the phase-field simulations show that $P_2$ becomes nonzero as well. The maximum values of $P_1$ and $P_2$ as a function of rotation angle $\theta$ are plotted in Figure 2 (b). The $P_1$ component has non-zero values at all angles, while the $P_2$ component is zero when $\theta = n\pi/4$ ($n$ is an arbitrary integer). In other words, the pure ferroelectric domain wall is Ising-Néel wall when $\theta = n\pi/4$ and is an Ising-Bloch-Néel wall for all other orientations.

As an example, Figure 3 (a) compares the domain wall profile of oblique walls with $\theta = \pi/12$ and $\theta = 5\pi/12$. Both $P_1$ and $P_2$ are nonzero near the domain walls, but vanish at the wall center. The profiles of all the polarization components are odd functions and they agree quite well with previous calculations [3,5,18] in magnitude and profile shape. Figure 3 (b) shows the result from first-principles calculations with $\theta \approx \pi/12$. We obtain this domain wall structure by expanding the dimensions of the single domain to $\sqrt{17}a \times \sqrt{17}a \times c$ and putting two such domains of opposite $P_3$ polarizations together along the $x_1$ direction. Thus the domain wall actually lies in the (410)$_C$ plane. Similar as the (100)$_C$ domain wall, we only consider the O-Ba-O plane as the centered wall. Also as we dealt with the (100)$_C$ domain wall, we relaxed this configuration with symmetry constraints and optimized the length of the normal-to-wall lattice vector. After the geometry optimization both $P_1$ and $P_2$ are non-zero. Again the polarization components from first-principles are larger than those calculated from phase-field method, but the peak positions are very close. The bound charge is calculated to be $2.3 \times 10^7$ C/m$^3$, indicating possible strong
9

interactions with ionic defects and the conductivity change. Qualitatively, both computational methods confirm that the 180° domain wall is Ising-Bloch-Néel-like when $\theta \neq n\pi/4$.

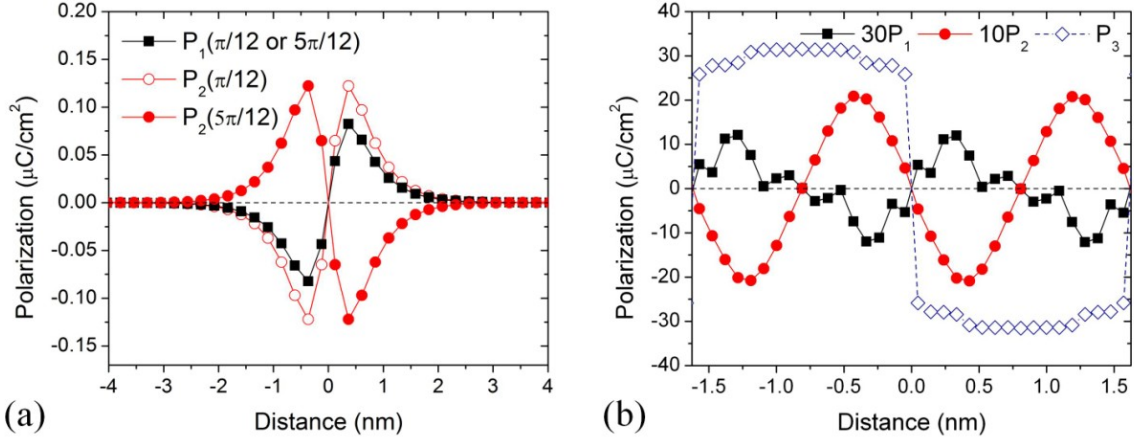

Figure 3. Polarization profiles of 180° domain walls. (a) $P_1$ and $P_2$ distribution of $\theta = 5\pi/12$ and $\theta = \pi/12$ from phase-field method. $P_1$ remains identical while $P_2$ flips with $\theta$. Both $P_1$ and $P_2$ are independent of $P_3$, so $P_3$ is neglected in this plot. (b) $P_1$ and $P_2$ profiles with $\theta = 5\pi/12$ from first-principles calculation.

From Landau-Ginzburg-Devonshire (LGD) analytical theory, we also derived an approximate expression of $P_1$, $P_1(x_1, \theta) \approx \varepsilon_0 \varepsilon_b P_S^2 \dfrac{2F_{12}}{R_c} \dfrac{Q_{11}+Q_{12}}{s_{11}+s_{12}} \dfrac{\sinh(x_1/R_c)}{\cosh^3(x_1/R_c)}$, where $R_c$ is the correlation radius, $F_{12} = F_{12}^C + \dfrac{1}{2}\sin^2(2\theta)(F_{11}^C - F_{12}^C - F_{44}^C)$, and $P_S$ is the spontaneous polarization far from the domain wall (see supplementary for details of derivations). The odd distribution of $P_1(x_1, \theta) \propto \dfrac{\sinh(x_1/R_c)}{\cosh^3(x_1/R_c)}$ is in agreement with the phase-field profile shown in Figure 2 (c). The angular dependence of the maximal value $P_1^{\max}(\theta) \propto F_{12}^C + \dfrac{1}{2}\sin^2(2\theta)(F_{11}^C - F_{12}^C - F_{44}^C)$ also agrees qualitatively with the polar plot (Figure 2 (b)). Due to the coupling with $P_1$, the analytical expression for $P_2$ is difficult to derive. With an artificial condition of $P_1 \equiv 0$, previous study [5] obtained that $P_2^{max}(\theta) \propto \sin(4\theta)(F_{44}^C - F_{11}^C + F_{12}^C)$. However, as shown in our phase-field results and first-principles calculations, $P_1$ and $P_2$ have similar magnitudes. Therefore $P_1$ should not be neglected, and this expression should be taken qualitatively.

The flexoelectric effect induced polarization components exhibit unique properties. An interesting feature among them is the chirality of $P_1$ and $P_2$ profiles as shown in Figure 3(a). Firstly, three polarization components seem to be independent of each other. This can be simply explained by the small magnitude of $P_1$ and $P_2$. Secondly, the profile of $P_2$ flips when the rotation angle goes from $\theta$ to its complementary angle $\pi/2-\theta$, while chirality of the Néel wall is apparently independent of rotation angle $\theta$. This means that the Néel wall is always tail-to-tail regardless of the wall orientation, which gives rise to a unique bound charge distribution pattern. We will demonstrate below that the chirality can be explained within the framework of LGD analytical theory by including the flexoelectric effect.

By minimizing the total free energy Equation (2), we get the equations of state (see supplementary for details of derivation)

$$g_{ijkl} \frac{\partial^2 P_k}{\partial x_j \partial x_l} = -E_i^f - E_i^d \qquad (3)$$

where $E_i^f = F_{ijkl} \frac{\partial \sigma_{kl}}{\partial x_j}$ is the so-called flexoelectric field [19], which is used to describe the flexoelectric effect, and $F_{ijkl}=s_{ijmn}f_{mnkl}$ is the flexoelectric field tensor [19]. Due to the electrostrictive effect, the stress distribution of the domain wall is always symmetric with respect to the domain wall center as illustrated in Figure 4 (a). The flexoelectric fields are thus odd functions since they are proportional to the stress gradient. This feature is not limited to the pure ferroelectric but applied to all domain walls. The depolarization and flexoelectric fields in $x_1$ and $x_2$ directions are plotted in Figure 4 (b). The flexoelectric field $E_1^f$ is around an order of magnitude larger than $E_2^f$ because the gradient of $\sigma_{33}$ is much larger than that of $\sigma_{22}$. However, it is greatly weakened by the depolarization field. This explains why larger flexoelectric field cannot induce larger $P_1$ as compared to $P_2$. The flexoelectric and depolarization fields at the domain wall are thus the key factors determining the magnitude of the flexoelectricity-induced Néel and Bloch type polarization components.

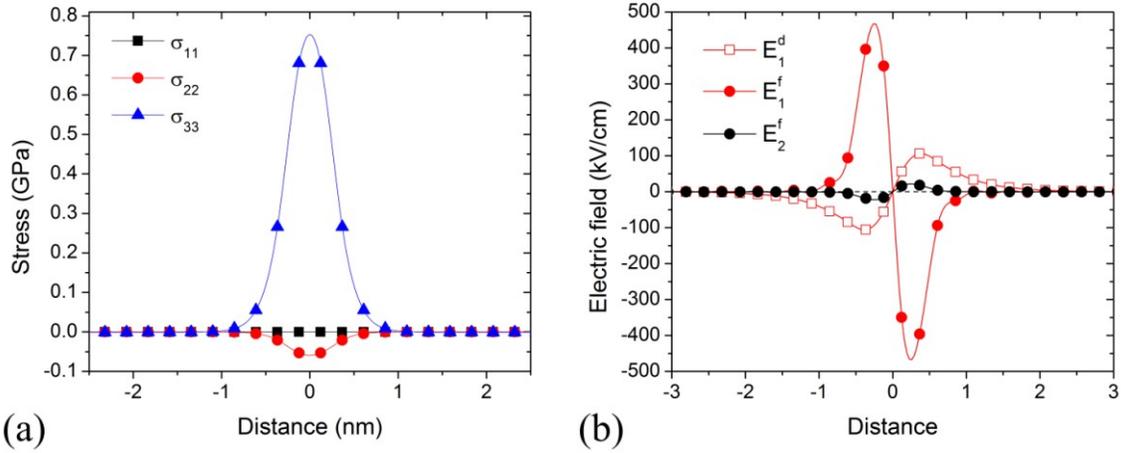

Figure 4. (a) Stress distribution around the domain wall with $\theta=5\pi/12$ from phase-field method, (b) driving forces of $P_2$ evolution around the domain wall with $\theta=5\pi/12$ from phase-field method. The lines are guide to the eye.

From the discussion above, we conclude that the $P_1$ component cannot be entirely screened. Furthermore, the existence of Néel feature is not limited to pure ferroelectric domain walls, i.e., it can be a general phenomenon that is also present in other kinds of domain walls. For example, in SrTiO$_3$ twin walls, the polarization is induced by the so-called *roto-flexo* field [20,21]. All four kinds of domain walls discussed in reference [20] exhibit nonzero $P_1$, which is greatly suppressed by the depolarization field in each case. This prediction also gets support from both resonant ultrasound spectroscopy [22] and resonant piezoelectric spectroscopy [23].

Since the depolarization field only exists in $x_1$ direction, one may expect larger $P_2$ or $P_3$ from the flexoeletric effect in certain domain wall configurations. With the flexoelectric coefficient $F$ of the order of $\sim 10^{-11}$ C$^{-1}$m$^3$, stress $\sigma$ at the wall of $\sim 1$ GPa, domain wall width of $\sim 1$ nm, the polarization induced by flexoelectric effect at the wall is estimated to be $\sim 1 \mu$C/cm$^2$. For example, in SrTiO$_3$, $P_3$ in the hard antiphase boundaries and $P_2$ in hard twins can reach as high as several $\mu$C/cm$^2$ due to the large gradients of $\sigma_{33}$ and $\sigma_{22}$. [20,21,24] The flexoelectric effect thus enables us a new way to control electric polarization rotation from domain wall engineering. By further manipulating the density, the domain wall can be used to tune piezoelectricity.

In conclusion, we have successfully developed a phase-field model of ferroelectric domains with flexoelectric effects. Both phase-field model and first-principles calculations predict that the classical Ising ferroelectric domain walls also possess Néel-like and Bloch-like features. We demonstrate that the additional components are produced by the flexoelectric effect. The additional polarization components are more than two magnitudes smaller than the Ising component, but it gives rise to a large bound charge density distribution. The chirality of Néel component is independent of domain wall orientation, while the Bloch chirality is not. Since the Néel component is induced by stress inhomogeneity at the domain walls, its existence is a general phenomenon, and its magnitude is determined by the competition between flexoelectric and depolarization effects.

Acknowledgement

YG would like to thank Dr. J. M. Hu for helpful suggestions. This work was supported by the NSF through grants DMR-0820404, DMR-1006541 and DMR-1210588. The computer simulations were carried out on the LION and cyberstar clusters at the Pennsylvania State University, in part supported by instrumentation (cyberstar Linux cluster) funded by the NSF through Grant OCI-0821527. ANM and EAE acknowledge the financial support via bilateral SFFR-NSF project (US National Science Foundation under NSF-DMR-1210588 and State Fund of Fundamental State Fund of Fundamental Research of Ukraine, grant UU48/002).

# Supplemental Material to

# Non-Ising Character of a Ferroelectric Wall Arises from a Flexoelectric Effect

Yijia Gu[1], Menglei Li[1,2], Anna N. Morozovska[3,4], Yi Wang[1], Eugene A. Eliseev[4], Venkatraman Gopalan[1], and Long-Qing Chen[1]

[1] Department of Materials Science and Engineering, Pennsylvania State University, University Park, PA 16802, USA

[2] Department of Physics and State Key Laboratory of Low-Dimensional Quantum Physics, Tsinghua University, Beijing 100084, People's Republic of China

[3] Institute of Physics, National Academy of Science of Ukraine, 46, pr. Nauki, 03028 Kiev, Ukraine

[4] Institute for Problems of Materials Science, National Academy of Science of Ukraine, Krjijanovskogo 3, 03142 Kiev, Ukraine

**This file includes:**

Supplementary Discussions 1-4

Supplementary References

Supplementary Figure

Supplementary Tables S1-3

# 1. Derivation and additional discussion on Equation (3)

By minimizing the total free energy Eq. (1) with respect to polarization, we get

$$g_{ijkl}\frac{\partial^2 P_k}{\partial x_j \partial x_l} = 2\alpha_{ij}P_j - Q_{ijkl}P_j\sigma_{kl} - F_{ijkl}\frac{\partial \sigma_{kl}}{\partial x_j} - E_i^d \quad (S.1)$$

where $F_{ijkl}=s_{ijmn}f_{mnkl}$ is the flexoelectric field coefficient [1]. (Higher order terms of $P$ polynomials are ignored for simplicity.) The first two terms on the right hand side are really small at the domain walls (see the second part of the supplementary for details). So the equation (S.1) can be further reduced to

$$g_{ijkl}\frac{\partial^2 P_k}{\partial x_j \partial x_l} = -E_i^f - E_i^d \quad (S.2)$$

where $E_i^f = F_{ijkl}\frac{\partial \sigma_{kl}}{\partial x_j}$ is the so-called flexoelectric field [1], which is used to describe the flexoelectric effect.

Equation (S.2) demonstrates that the polarization is essentially governed by the competition between flexoelectric field and the depolarization field. With the stress-free boundary condition, we have $\sigma_1= \sigma_5= \sigma_6= 0$ in the wall and

$$\frac{\partial^2 P_1}{\partial x_1^2} = \frac{(E_1^d + E_1^f)g_{66} - E_2^f g_{16}}{g_{16}^2 - g_{11}g_{66}} \quad (S.3a)$$

$$\frac{\partial^2 P_2}{\partial x_1^2} = \frac{E_2^f g_{11} - (E_1^d + E_1^f)g_{16}}{g_{16}^2 - g_{11}g_{66}} \quad (S.3b)$$

where $g_{11} = g_{11}^C + \frac{g_A^C(\cos 4\theta - 1)}{4}$, $g_{16} = -\frac{g_A^C \sin 4\theta}{4}$, $g_{66} = g_{44}^C - \frac{g_A^C(\cos 4\theta - 1)}{4}$,

$E_1^f = F_{12}^C(\frac{\partial \sigma_3}{\partial x_1} + \frac{\partial \sigma_2}{\partial x_1}) - \frac{F_A^C(\cos 4\theta - 1)}{4}\frac{\partial \sigma_3}{\partial x_1}$, $E_2^f = \frac{F_A^C \sin 4\theta}{4}\frac{\partial \sigma_2}{\partial x_1}$, $g_A^C = g_{11}^C - g_{12}^C - 2g_{44}^C$ and

$F_A^C = F_{11}^C - F_{12}^C - F_{44}^C$. The superscript "C" denotes the tensors in the original crystallographic coordinate of pseudocubic lattice. The indices are simplified following Voigt notation. The tensors in the rotated coordinate system as functions of domain wall angle $\theta$ are listed in **Table S1**.

The stress distribution of the domain wall is always symmetric with respect to the domain wall center as illustrated in Figure 4 (a). The flexoelectric fields are thus odd functions since they are proportional to the stress gradient. Then from Eq. (S.3), we notice that the curvature of $P_1$ is a function of $\cos(4\theta)$ while for $P_2$ it is a function of $\sin(4\theta)$. This exactly explains the symmetry that we observed in the polar plot of Figure 2(d) and the chirality change of $P_2$ component with respect to rotation angle $\theta$. Eq. (S.3a) also indicates that the existence of $P_1$ does not depend on the background dielectric constant but rather the flexoelectric field.

## 2. Derivation of the analytical expressions for $P_1$

Within Landau-Ginzburg-Devonshire theory, equations of state for polarization components depending only on $x_1$ have the form [2]:

$$2a_1 P_1 + 4a_{11} P_1^3 + 2a_{12} P_2^2 P_1 + 2a_{12} P_3^2 P_1 + a_{16} P_2 \left(3P_1^2 - P_2^2\right) - $$
$$-g_{11}\frac{\partial^2 P_1}{\partial x_1^2} - g_{16}\frac{\partial^2 P_2}{\partial x_1^2} - 2(Q_{12}\sigma_3 + Q_{12}\sigma_2)P_1 - Q_{26}\sigma_2 P_2 = E_1^d + F_{12}\left(\frac{\partial \sigma_2}{\partial x_1} + \frac{\partial \sigma_3}{\partial x_1}\right) \quad \text{(S.4a)}$$

$$2a_1 P_2 + 4a_{11} P_2^3 + 2a_{12} P_1^2 P_2 + 2a_{12} P_3^2 P_2 + a_{16} P_1\left(P_1^2 - 3P_2^2\right) -$$
$$-g_{66}\frac{\partial^2 P_2}{\partial x_1^2} - g_{16}\frac{\partial^2 P_1}{\partial x_1^2} - 2(Q_{12}\sigma_3 + Q_{11}\sigma_2)P_2 - Q_{26}\sigma_2 P_1 = F_{26}\frac{\partial \sigma_2}{\partial x_1} \quad \text{(S.4b)}$$

$$2a_1 P_3 + 4a_{11} P_3^3 + 2a_{12}\left(P_1^2 + P_2^2\right) P_3 - g_{44}\frac{\partial^2 P_3}{\partial x_1^2} - 2\left(Q_{11}\sigma_3 + Q_{12}\sigma_2\right)P_3 - Q_{44}\sigma_4 P_2 = 0 \quad (S.4c)$$

The right-hand-side of Eq. (S.4a) can be written as

$$F_{12}\frac{\partial(\sigma_2 + \sigma_3)}{\partial x_1} \approx \frac{F_{12}(Q_{11} + Q_{12})}{s_{11} + s_{12}}\frac{\partial\left(-P_2^2 - P_3^2\right)}{\partial x_1} \quad (S.5)$$

Elastic stresses are:

$$\sigma_1 = \sigma_5 = \sigma_6 = 0, \quad \sigma_2 = \frac{s_{11} U_2 - s_{12} U_3}{s_{11}^2 - s_{12}^2}, \quad \sigma_3 = \frac{s_{11} U_3 - s_{12} U_2}{s_{11}^2 - s_{12}^2}, \quad \sigma_4 = \frac{Q_{44}\left(P_2^S P_3^S - P_2 P_3\right)}{s_{44}} \quad (S.6)$$

where $P_i^S$ is the spontaneous polarization component in $x_i$ direction. The tensors in rotated coordinate system are listed in the **Table S1.** Functions $U_3$ and $U_2$ are defined as

$$U_3 = Q_{11}\left(\left(P_3^S\right)^2 - P_3^2\right) + Q_{12}\left(\left(P_2^S\right)^2 + \left(P_1^S\right)^2 - \left(P_2^2 + P_1^2\right)\right) + F_{12}\frac{\partial P_1}{\partial x_1}, \quad (S.7a)$$

$$U_2 = Q_{11}\left(\left(P_2^S\right)^2 - P_2^2\right) + Q_{12}\left(\left(P_3^S\right)^2 + \left(P_1^S\right)^2 - \left(P_1^2 + P_3^2\right)\right) + F_{12}\frac{\partial P_1}{\partial x_1}, \quad (S.7b)$$

Thus one obtains

$$\sigma_2 + \sigma_3 = \frac{U_2 + U_3}{s_{11} + s_{12}} \equiv \frac{Q_{11} + Q_{12}}{s_{11} + s_{12}}\left(P_S^2 - P^2\right) + \frac{Q_{12} - Q_{11}}{s_{11} + s_{12}}\left(\left(P_1^S\right)^2 - P_1^2\right) + \frac{2F_{12}}{s_{11} + s_{12}}\frac{\partial P_1}{\partial x_1} \quad (S.8)$$

where $P_S^2 = \left(P_1^S\right)^2 + \left(P_2^S\right)^2 + \left(P_3^S\right)^2$ and $P^2 = P_1^2 + P_2^2 + P_3^2$. With the inequalities of polarization components $|P_2| \ll |P_3|$ and $|P_1| \ll |P_3|$, we can rewrite Eq.(S.4a) as

$$\left(\frac{1}{\varepsilon_0\varepsilon_b} + 2a_1 + 2a_{12} P_3^2 - 2Q_{12}\frac{Q_{11} + Q_{12}}{s_{11} + s_{12}}\left(P_S^2 - P^2\right)\right) P_1 + 4a_{11} P_1^3$$

$$-\left(g_{11} + \frac{2F_{12}^2}{s_{11} + s_{12}}\right)\frac{\partial^2 P_1}{\partial x_1^2} \approx -F_{12}\frac{Q_{11} + Q_{12}}{s_{11} + s_{12}}\frac{\partial P_3^2}{\partial x_1} \quad (S.9)$$

The factor $\frac{1}{\varepsilon_0 \varepsilon_b}$ comes from the depolarization field. With parameters from the **Table S2**, we can simplify the linear and gradient terms in (S.9) as

$$\frac{1}{\varepsilon_0 \varepsilon_b} + 2a_1 + 2a_{12}P_3^2 - 2Q_{12}\frac{Q_{11}+Q_{12}}{s_{11}+s_{12}}\left(P_S^2 - P^2\right) \approx \frac{1}{\varepsilon_0 \varepsilon_b} \qquad (S.10a)$$

Because in magnitude $2Q_{12}\frac{Q_{11}+Q_{12}}{s_{11}+s_{12}}\left(P_S^2 - P^2\right) \approx 2a_{12}P_3^2 \approx 2a_1 = -6\times 10^7$ mJ/C$^2 \cdot$ and $\frac{1}{\varepsilon_0 \varepsilon_b} \approx$ $1.6\times 10^{10}$ m/F, the contribution induced by depolarization field is more than 100 times larger. So the ferroelectric nonlinearity term in Eq. (S.4a) can be neglected as well.

Far away from the domain wall, the derivative can be estimated as

$$\left(g_{11} + \frac{2F_{12}^2}{s_{11}+s_{12}}\right)\frac{\partial^2 P_1}{\partial x_1^2} \approx \left(g_{11} + \frac{2F_{12}^2}{s_{11}+s_{12}}\right)\frac{P_1}{R_c^2} < 10^9 P_1 \ll \frac{P_1}{\varepsilon_0 \varepsilon_b} \qquad (S.10b)$$

where $R_c \geq 0.5\times 10^{-9}$m is the correlation radius. Since $\frac{2F_{12}^2}{s_{11}+s_{12}}$ has the similar magnitude as $g_{11} = 5\times 10^{-10}$C$^{-2}$m$^3$J, the gradient term is at least 10 times smaller than $\frac{P_1}{\varepsilon_0 \varepsilon_b}$. Thus it can be omitted as well. Eventually without losing accuracy, Eq. (S.9) can be simplified as $\frac{P_1}{\varepsilon_0 \varepsilon_b} \approx -F_{12}\frac{Q_{11}+Q_{12}}{s_{11}+s_{12}}\frac{\partial P_3^2}{\partial x_1}$. Thus the approximate expression for $P_1$ is

$$P_1 \approx -\varepsilon_0 \varepsilon_b F_{12}\frac{Q_{11}+Q_{12}}{s_{11}+s_{12}}\frac{\partial P_3^2}{\partial x_1} \qquad (S.11)$$

Using the approximation $P_3 \approx P_S \tanh(x_1/R_c)$ and the strong inequality $|P_2| \ll |P_3|$, we obtained that

$$P_1 \approx 2\varepsilon_0\varepsilon_b F_{12} \frac{Q_{11}+Q_{12}}{s_{11}+s_{12}} \frac{\sinh(x_1/R_c)}{\cosh^3(x_1/R_c)}$$

$$\equiv \frac{\varepsilon_0\varepsilon_b}{R_c} P_S^2 \left(2F_{12}^C - \sin^2(2\theta)\left(F_{44}^C + F_{12}^C - F_{11}^C\right)\right) \frac{Q_{11}+Q_{12}}{s_{11}+s_{12}} \frac{\sinh(x_1/R_c)}{\cosh^3(x_1/R_c)} \tag{S.12}$$

## 3. Discussion on the sensitivity of induced polarization component on various parameters

To discuss the sensitivity of the calculated polarization on the parameters used in the phase-field simulations, we divide the parameters into three groups: the parameters related to the thermodynamic potential, the flexoelectric coupling coefficients and the dielectric constant.

i) The parameters related to the thermodynamic potential include the Landau-Devonshire coefficients, the elastic stiffness constants, gradient energy coefficients, electrostrictive constants. Actually there are at least four sets of parameters available for the thermodynamic potential of $BaTiO_3$. [3–6] In our phase-field simulation, we chose the parameters from J. J. Wang et al [3]. As demonstrated in his work, this set of parameters can reproduce the spontaneous polarization, dielectric constant, temperature-electric field phase diagram and piezoelectric coefficients. In addition, this set of parameters was shown to be the best in reproducing the electric field induced tetragonal to orthorhombic ferroelectric transition [7].

ii) The flexoelectric coefficients from experiment measurement are three orders of magnitude larger than the calculated values [8]. The disagreement may come from different boundary conditions used [9], surface flexoelectric effect, dynamic flexoelectric effect and etc [8]. It is more reasonable to use the calculated values from first-principles. In order to evaluate the sensitivity of the calculated results on the values of flexoelectric coefficients we made a comparison using two sets of flexoelectric coefficients (**Table S3**) [10,11]. Although, both sets of flexoelectric coefficients are from first-principles calculations, they are quite different in terms of magnitude and sign. As shown in Figure S1 (a-d), although the flexoelectric effect induced polarization components, namely $P_1$ and $P_2$, are dependent on the flexoelectric coefficients in terms of profiles and anisotropy, they still exist. The magnitudes of both induced polarization components do not change much despite the huge difference between the two sets of flexoelectric coefficients. The profile of $P_1$ clearly flipped, which is mainly due to the sign change of $f_{12}$.

iii) As demonstrated in Equations (S2) and (S3), the depolarization field $E^d$ has strong effect on the induced polarization. Therefore, the background dielectric constant, which determines the strength of the electric field, is another important parameter. In order to discuss the sensitivity of induced polarization on background dielectric constant, we need to start from the two contributions to the polarization of a ferroelectric material: (1) the critical displacements of ions (responsible for enhanced dielectric constant and spontaneous polarization); (2) all the other polar distortions. The background dielectric constant is from the latter contribution [12]. So we chose two quite different values as listed in Table S3. One is 7.35, which is essentially from room temperature infrared and Raman reflectivity data [13]. The other one is 45, taken from Rupprecht and Bell's work [14]. As pointed out by Ref. [14], the background dielectric constant "consists of contributions from the electric polarizability, temperature-independent optically active lattice vibrations, and a dominant term stemming from the finite frequency of the temperature-dependent soft mode in the limit of infinite temperature". Thus the value 7.35, which is from the optical modes only, is not sufficient. However, the simulation results do not show much difference as we compare Figure 1 (c, d) and (e, f). The Néel and Bloch features still exist and maintain their anisotropy. Only the magnitudes of both components are about two times larger than our previous calculations. That's because the increased background dielectric constant weakened the depolarization field.

From the above analysis, we can conclude that the induced polarization components calculated from phase-field simulation do depend on the parameters. Their anisotropy is dependent on the flexoelectric coefficients. And their magnitude is dependent on both the flexoelectric coefficients and the background dielectric constant. But physical phenomenon, flexoelectric effect induced two new polarization components at the 180˚ domain wall, is not sensitive to the parameters. As demonstrated by our analysis, the Néel-like ($P_1$) and Bloch-like ($P_2$) polarization components are essentially dominated by the Equitation (S3). Both the flexoelectric field and the depolarization field are functions of polarization. Therefore, only in very some special cases, for example $F_I^C = g_I^C = g_{44}^C = 0$, there are no $P_1$ and $P_2$ with any domain wall orientation. To prove the reliability of our calculations, we compares our simulation result of BaTiO$_3$ with Lee's DFT calculations of PbTiO$_3$ [15] as shown in Figure S2. Although the materials and the computational methods are quite different, the polarization profiles are very similar.

## 4. Computational details for first-principles calculation

We performed first-principles calculation using density functional theory as implemented in the Vienna *ab-initio* simulation package [16]. We employed the projector augmented wave method [17,18] with an energy cutoff of 400 eV and generalized gradient approximations with Perdew-Burke-Ernzerhof pseudopotentials revised for solids (PBEsol) [19]. Ba $5s$, $5p$ and $6s$ electrons, Ti $3s$, $3p$, $3d$ and $4s$ electrons and O $2s$ and $2p$ electrons were treated as valence states. We relaxed the atom positions along with the length of the lattice vector normal to the domain wall plane using the conjugate gradient algorithm [20] until the residual forces were smaller than 0.01eV/Å. The Born effective charge tensors were calculated with density functional perturbation theory. A 1×9×9 Monkhorst-Pack k-mesh is used for $(100)_C$ domain walls and a 1×3×7 k-mesh is used for $(140)_C$ domain wall. The accuracy has been checked to be sufficient.

## Supplementary references

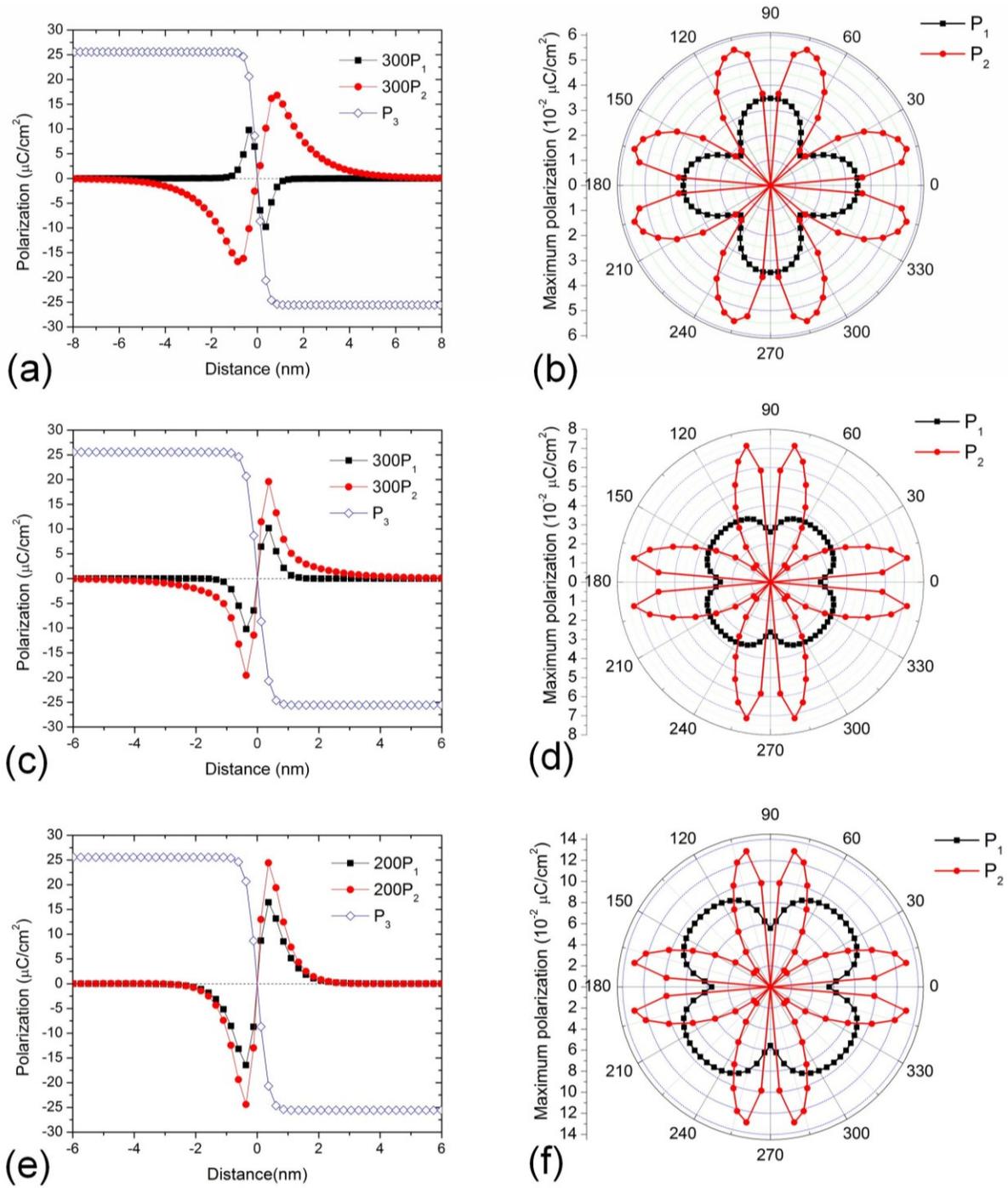

**Figure S1**. Polarization profiles with $\theta = \pi/12$ and the maximum values of $P_1$ and $P_2$ components as a function of domain wall orientation ($\theta$ is the angle between the domain wall and the [010] direction). (a) and (b) are calculated with flexoelectric coefficients set #1 and background dielectric constant of 7.35. (c)

and (d) are calculated with flexoelectric coefficients set #2 and background dielectric constant of 7.35. (e) and (f) are calculated with flexoelectric coefficients set #2 and background dielectric constant of 45.

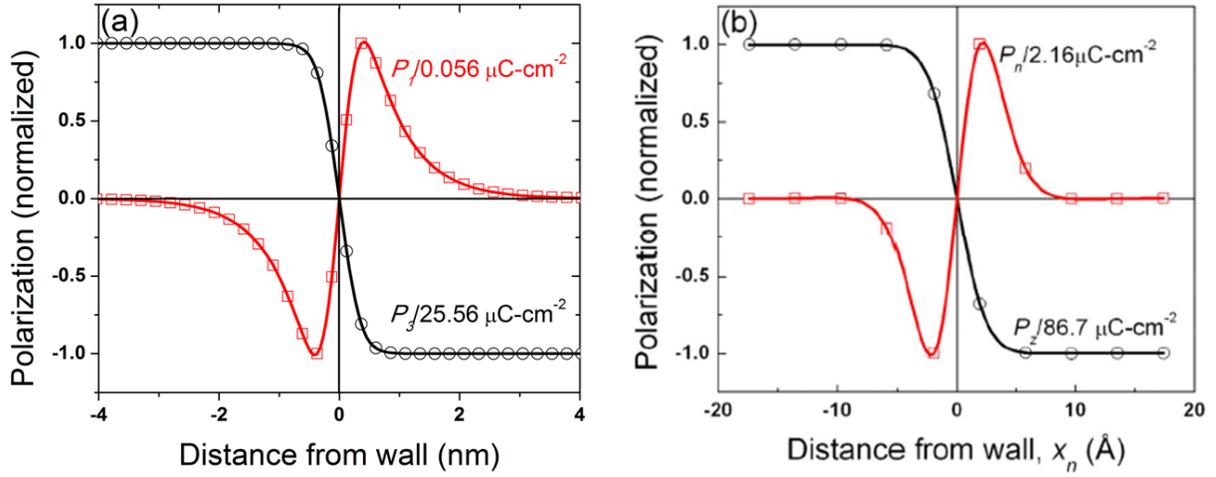

**Figure S2**. Domain wall profiles of (a) BaTiO$_3$ calculated from phase-field simulation (this work) and (b) PbTiO$_3$ calculated from first-principles (D. Lee et al [15]). P$_1$ in (a) corresponds to P$_n$ (normal) in (b), and P$_3$ in (a) corresponds to P$_z$ in (b).

**Table S1.** Dependence of the tensors and other coefficients on the wall rotation angle $\theta$ in the tetragonal phase (adapted from Ref. [21]).

| | |
|---|---|
| Elastic compliance components $s_{ij}$ in rotated coordinate system $\{x_1, x_2, x_3\}$ | $s_{11} = s_{11}^C - \dfrac{\sin^2(2\theta)}{2}s_A^C$, $s_{12} = s_{12}^C + \dfrac{\sin^2(2\theta)}{2}s_A^C$, $s_{16} = -s_{26} = -\dfrac{\sin(4\theta)}{2}s_A^C$, $s_{66} = s_{44}^C + 2\sin^2(2\theta)s_A^C$, with $s_A^C = s_{11}^C - s_{12}^C - \dfrac{s_{44}^C}{2}$ |
| Electrostriction tensor components $Q_{ij}$ in rotated coordinate system $\{x_1, x_2, x_3\}$ | $Q_{11} = Q_{11}^C - \dfrac{\sin^2(2\theta)}{2}Q_A^C$, $Q_{12} = Q_{12}^C + \dfrac{\sin^2(2\theta)}{2}Q_A^C$, $Q_{16} = -Q_{26} = -\dfrac{\sin(4\theta)}{2}Q_A^C$, $Q_{66} = Q_{44}^C + 2\sin^2(2\theta)Q_A^C$, with $Q_A^C = Q_{11}^C - Q_{12}^C - \dfrac{Q_{44}^C}{2}$ |
| Flexoelectric field tensor components $F_{ij}$ in rotated coordinate system $\{x_1, x_2, x_3\}$ | $F_{11} = F_{11}^C - \dfrac{1}{2}\sin^2(2\theta)F_A^C$, $F_{12} = F_{12}^C + \dfrac{1}{2}\sin^2(2\theta)F_A^C$, $F_{16} = -F_{26} = -\dfrac{\sin(4\theta)}{4}F_A^C$, $F_{66} = F_{44}^C + \sin^2(2\theta)F_A^C$, $F_{61} = 2F_{16}$, $F_{62} = 2F_{26}$, with $F_A^C = F_{11}^C - F_{12}^C - F_{44}^C$ |
| Gradient energy coefficients $g_{ij}$ in the rotated coordinate system $\{x_1, x_2, x_3\}$ | $g_{11} = g_{11}^C - \dfrac{\sin^2(2\theta)}{2}g_A^C$, $g_{16} = -\dfrac{\sin(4\theta)}{4}g_A^C$, $g_{66} = g_{44} + \dfrac{\sin^2(2\theta)}{2}g_A^C$, with $g_A^C = g_{11}^C - g_{12}^C - 2g_{44}^C$ |
| Landau-Devonshire coefficients $a_{ij}$ in the rotated coordinate system $\{x_1, x_2, x_3\}$ | $a_{11} = a_{11}^C - \dfrac{2a_{11}^C - a_{12}^C}{4}\sin^2(2\theta)$, $a_{12} = a_{12}^C + 3\dfrac{2a_{11}^C - a_{12}^C}{2}\sin^2(2\theta)$, $a_{16} = \dfrac{2a_{11}^C - a_{12}^C}{2}\sin(4\theta)$ |

Subscripts 1, 2 and 3 denote Cartesian coordinates $x$, $y$, $z$ and Voigt's (matrix) notations are used: $a_{11} \equiv a_1$, $a_{1111} \equiv a_{11}$, $6a_{1122} \equiv a_{12}$, $g_{1111} \equiv g_{11}$, $g_{1122} \equiv g_{12}$, $g_{1212} \equiv g_{66}$, $Q_{1111} \equiv Q_{11}$, $Q_{1122} \equiv Q_{12}$, $4Q_{1212} \equiv Q_{44}$, $s_{1111} \equiv s_{11}$, $s_{1122} \equiv s_{12}$, $4s_{1212} \equiv s_{44}$, $F_{1111} \equiv F_{11}$, $F_{1122} \equiv F_{12}$, $2F_{1212} \equiv F_{44}$. Note that different factors (either "4", "2" or "1") in the definition of matrix notations with indices "44" are determined by the internal symmetry of tensors as well as by the symmetry of the corresponding physical properties tensors (see e.g. [22]).

**Table S2.** Material parameters of BaTiO$_3$

| Coefficients | Values (collected and recalculated mainly from Ref. [3]) |
|---|---|
| $\varepsilon_b$, $\varepsilon_0$ | $\varepsilon_b$=45 (Ref. [14]), $\varepsilon_0$=8.85×10$^{-12}$ F/m |
| $a_i$ (C$^{-2}$·mJ) | $a_1^C = 5\times10^5 T_S \left( \text{Coth}(\frac{T_S}{T}) - \text{Coth}(\frac{T_S}{390}) \right)$, $T_S$=160 K (at 293 K $a_1^C = -4.277\times10^7$) |
| $a_{ij}$ (×10$^8$ C$^{-4}$·m$^5$J) | $a_{11}^C = -1.154$, $a_{11}^C = 6.53$ |
| $a_{ijk}$ (×10$^9$ C$^{-6}$·m$^9$J) | $a_{111}^C = -2.106$, $a_{112}^C = 4.091$, $a_{123}^C = -6.688$ |
| $a_{ijkl}$ (×10$^{10}$ C$^{-6}$·m$^9$J) | $a_{1111}^C = 7.59$, $a_{1112}^C = -2.193$, $a_{1122}^C = -2.221$, $a_{1123}^C = 2.416$ |
| $Q_{ij}$ (C$^{-2}$·m$^4$) | $Q_{11}^C = 0.11$, $Q_{12}^C = -0.045$, $Q_{44}^C = 0.059$ |
| $s_{ij}$ (×10$^{-12}$ Pa$^{-1}$) | $s_{11}^C = 9.07$, $s_{12}^C = -3.19$, $s_{44}^C = 8.2$ |
| $g_{ij}$ (×10$^{-10}$C$^{-2}$m$^3$J) | $g_{11}^C = 5.1$, $g_{12}^C = -0.2$, $g_{44}^C = 0.2$ [23] |
| $F_{ij}$ (×10$^{-11}$C$^{-1}$m$^3$) | ~100 (estimated from measurements of Ref. [24]) $F_{11}^C = 0.3094$, $F_{12}^C = -0.279$, $F_{44}^C = -0.1335$ (recalculated from [11]) |

**Table S3.** The flexoelectric coefficients

| | $f_{11}$ | $f_{12}$ | $f_{44}$ | Ref. |
|---|---|---|---|---|
| Set #1 (Ba$_{0.5}$Sr$_{0.5}$TiO$_3$) | 59.86 nC/m | 38.81 nC/m | 0.526 nC/m | [10] |
| Set #2 (BaTiO$_3$, used in the main text) | 0.150 nC/m | -5.463 nC/m | -1.904 nC/m | [11] |